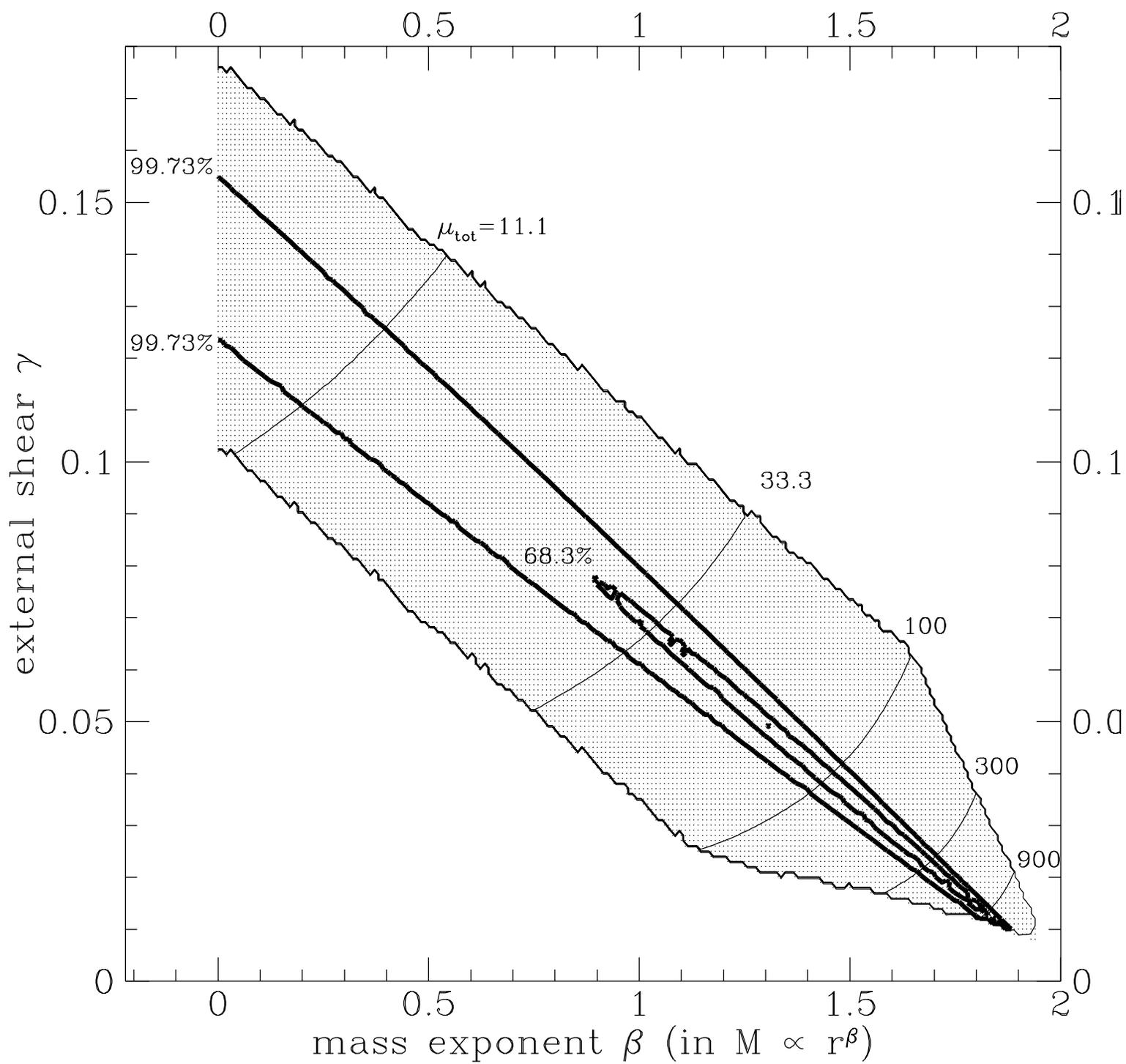

Figure 1

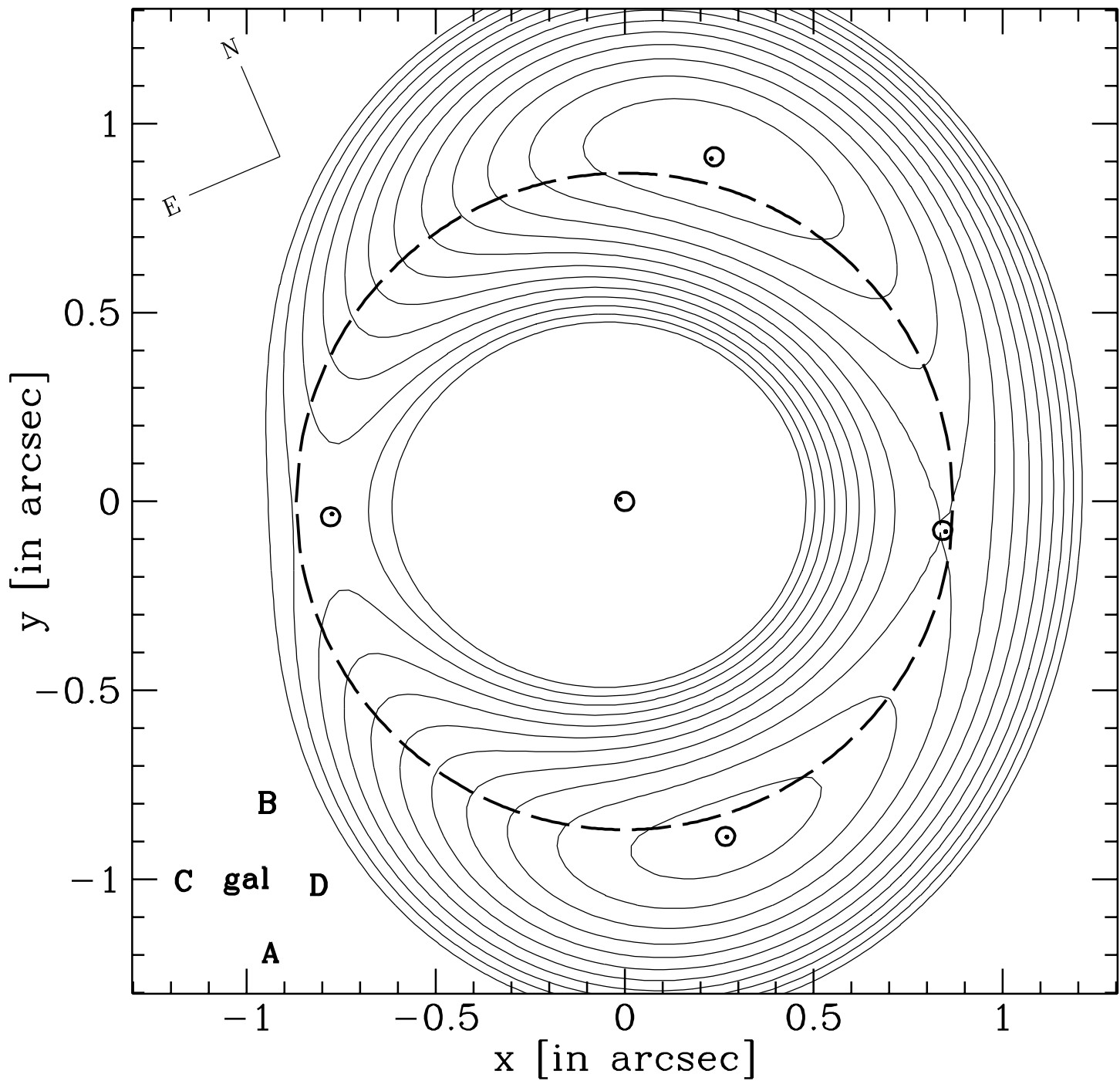

Figure 2

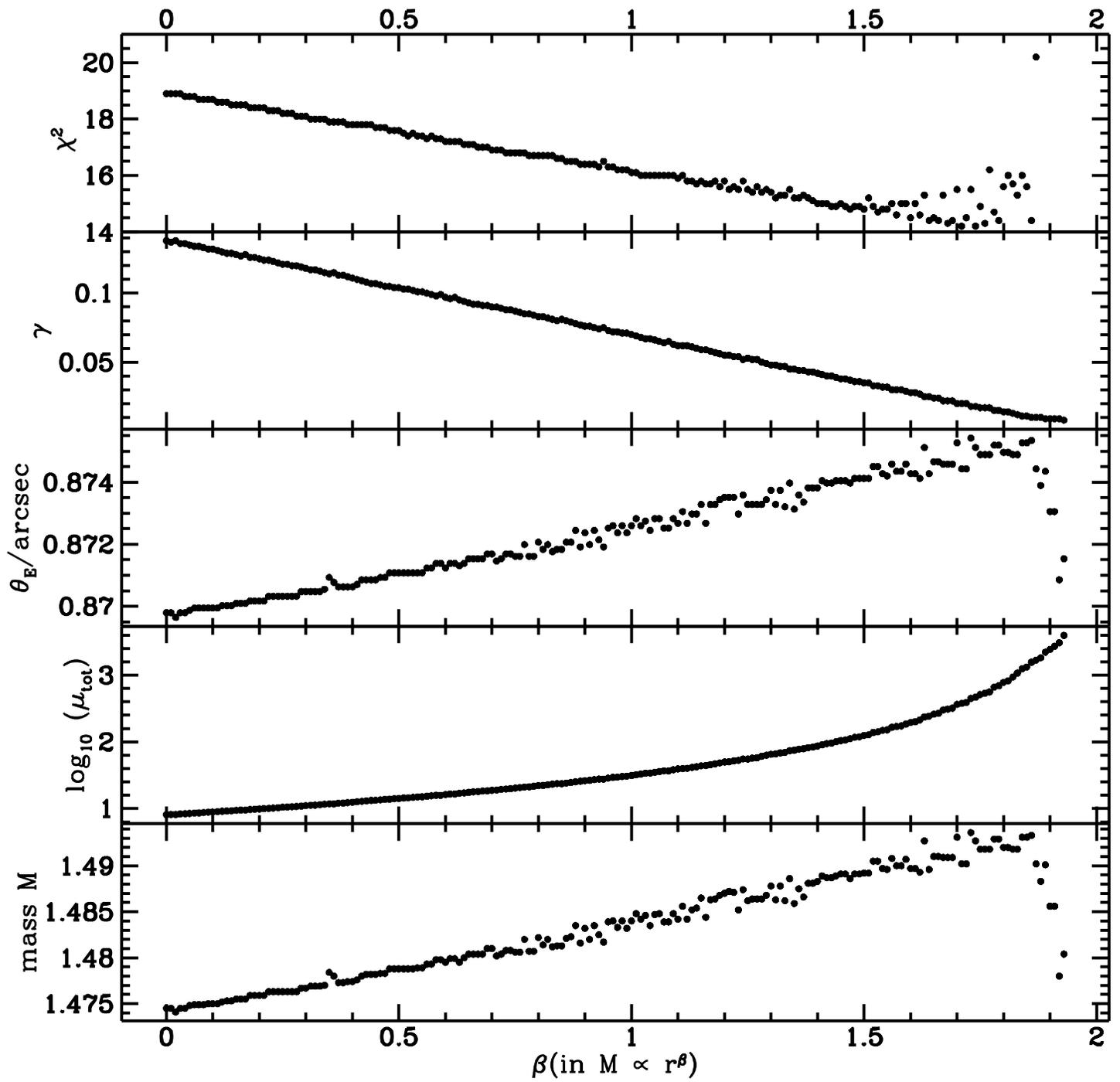

Figure 3

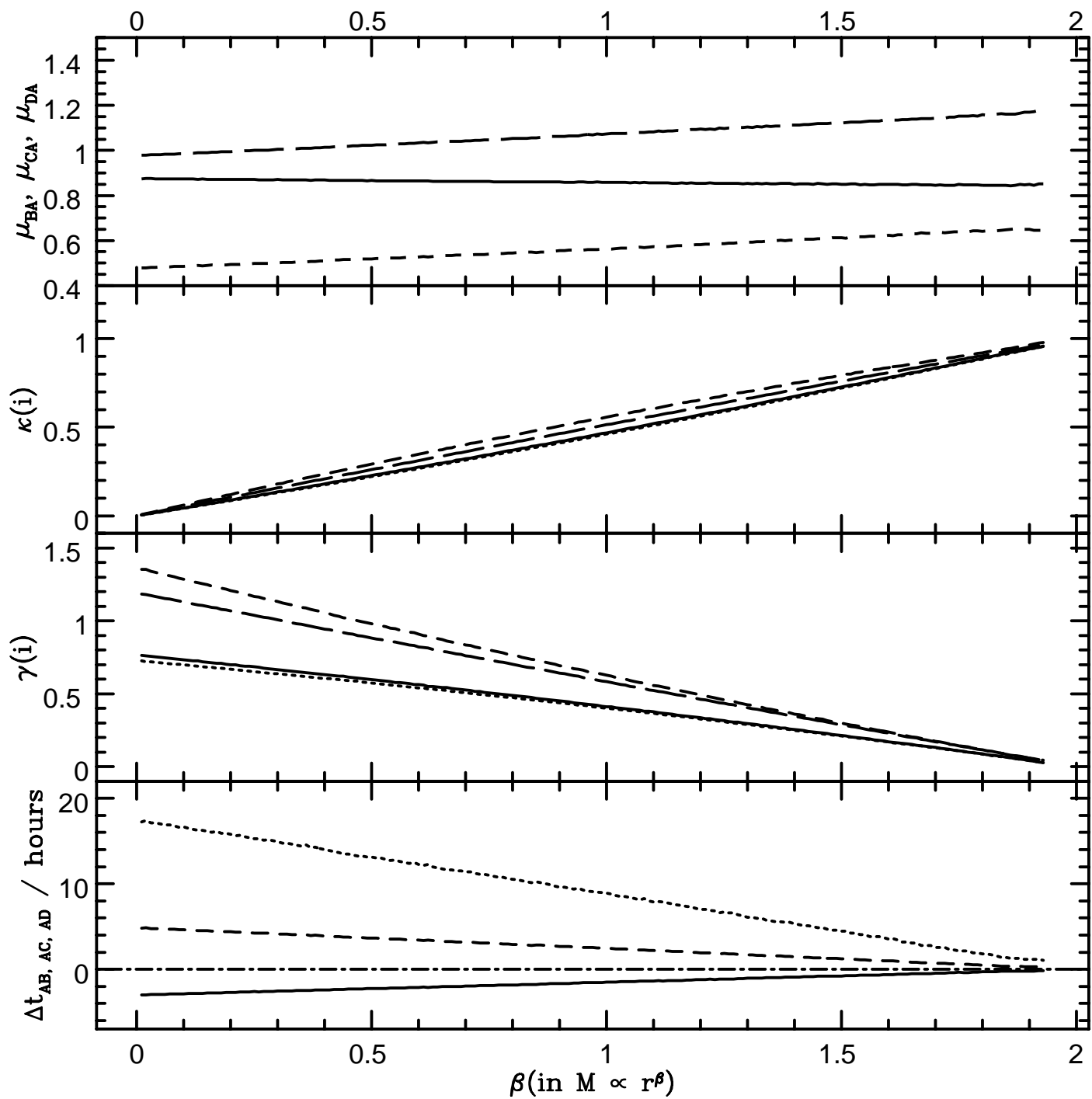

**Figure 4**

# Parameter degeneracy in models of the quadruple lens system Q2237+0305

Joachim Wambsganss[1,2] and Bohdan Paczyński[2]

[1] Max-Planck-Institut für Astrophysik, Karl-Schwarzschildstr.1, 85740 Garching, Germany

[2] Princeton University Observatory, Peyton Hall, Princeton, NJ 08544

## ABSTRACT

The geometry of the quadruple lens system Q2237+0305 is modeled with a simple astigmatic lens: a power-law mass distribution: $m \sim r^\beta$ with an external shear $\gamma$. The image positions can be reproduced with an accuracy better than 0.01 arcsec for any $0.0 \leq \beta \leq 1.85$ and the corresponding value of $\gamma = 0.1385 - 0.0689\beta$. This is a factor of $\sim 4$ more precise than what can be achieved by the best constant $M/L$ lens models (Rix et al. 1992). The image intensity ratios and the time delay ratios are almost constant along our one parameter family of models, but the total magnification varies from 8 to > 1000, and the maximum time delay (between leading image B and trailing image C) for $H_0$ of 75 km sec$^{-1}$ Mpc$^{-1}$ varies from more than 20 hours to about 1.5 hours, while $\beta$ increases from 0.0 to 1.85.

*Subject headings:* Gravitational lensing — quasars: general — quasars: individual: Q2237+0305 — numerical methods

## 1. Introduction

The quasar Q2237+0305 ($z_q = 1.695$) was found by Huchra et al. (1985) to be superimposed on the center of a barred spiral galaxy ($z_{gal} = 0.039$) and later identified as a gravitational lens system with four images (Tyson & Gorenstein 1985; Yee 1988; Schneider et al. 1988, subsequently S88). Lens models for this quadruple system were developed by S88 and Kent & Falco (1988) (from now on KF88): the theoretical and the observed positions differed by less than 0.2 arcsec (S88) and a few hundredths of an arcsec (KF88), respectively. The relative magnifications of the four images were reproduced to within a factor 1.5. KF88 used de Vaucouleurs and King profile mass distributions, whereas S88 assumed a constant mass-to-light ratio for the galaxy. Kochanek (1991, subsequently K91) used point lens and isothermal sphere mass distributions with different realizations of the quadrupole term in his models for the Q2237+0305. He finds for his models that the mass determination within the four images is insensitive to the mass distribution inside.

With positional accuracy improved by an order of magnitude by HST observations, Rix et al. (1992) (henceforth R92) improved the S88 work. Their model yielded image positions that deviated by about 0.03 arcsec from the observed positions. Although this is only a fraction of 2.5% of the image separation, it is still considerably larger than the positional error. R92 found furthermore that the mass of the galaxy inside a radius of 0.9 arcsec around the core is $m(r \leq 0.9'') \approx 1.08 \times 10^{10} M_\odot h_{100}^{-1}$ (within 2%).

According to the models of KF88 and S88/R92 the total magnification of all four quasar images is between 8 and 23, K91 finds values between 8 and 32. It was not clear from these studies, how unique the results were, and if models with higher total magnifications could be excluded. We started with the question: is it possible to construct simple lens models for Q2237+0305 with total magnifications very different from those obtained by earlier modelers? This has interesting aspects: if the lensing magnifications are one or two orders of magnitude higher than assumed, then the consequences from observed microlensing events on the size of the accretion disk around the quasar and the conclusions concerning the physics of quasars are strongly influenced (Rauch & Blandford 1991; Jaroszynski, Paczyński & Wambsganss 1992). A related question is: how reliable are the values of the surface mass density $\kappa_i$ and the local shear $\gamma_i$ for the four images? This is important for microlensing calculations, because the frequency and amplitude of high magnification events depend strongly on these values. Webster et al. (1991) have already noted that the microlensing parameters in this system are not very accurately determined, and that the published values (S88) are not unique. Q2237+0305 is the first gravitational lens in which microlensing has been discovered (Irwin et al. 1989, Corrigan et al. 1991). This lens system is unusual compared with all the other known lenses in that the lensing galaxy is so nearby. This makes it both very favorable for short microlensing time scales and for a detailed study of the lensing galaxy.

Two groups have recently obtained good positions for the Q2237+0305 system. Racine (1991) used CFHT data to determine the image positions and claims accuracy to ± 0."002. Crane et al. (1991) used HST data and found relative positions of the four images with formals errors of ± 0."005. All our models described below are based on the positions by Crane et al. (1991). However, we have run our program on the Racine (1991) data as well, and the results are qualitatively the same, although there are slight quantitative differences.

We do not attempt to construct a well motivated physical model for the galaxy. We rather ask: how well can the observed image properties be reproduced with a very simple, astigmatic lens model. We do not intend to present "better" or "more realistic" models, but rather to explore the issue of uniqueness. We find that many observed parameters are degenerate, as a one parameter family of solutions fits the data very well. Most of what we present here was done before in K91 for two sets of lens models and applied to the lens systems known at that time. We concentrate here on one gravitational lens system only, the best studied case of 2237+0305. We use new and more accurate positions, and we consider a one parameter family of lens models. We emphasize in particular the effect of the degeneracy on the time delay and its consequences on the Hubble constant. A more general approach to degeneracies in gravitational lens models was presented by Gorenstein, Falco and Shapiro (1988); they show that a number of invariance transformations leaves the observables of a lens system unchanged.

Our method is described in Section 2, the results are shown in Section 3, and the discussion and conclusions, in particular on the possibility of breaking the parameter degeneracy, are given in Section 4.

## 2. Method

As observational constraints we use the eight coordinates of the four Q2237+0305 images plus the two coordinates of the core of the galaxy: a total of ten parameters to fit. We do not use the relative magnifications of the quasar images as a constraint, because they are subject to large changes due to microlensing (see Corrigan et al. 1991, Houde & Racine 1994), and the observing baseline is too short to get the "real" ratios just from averaging (cf. Webster et al. 1991). Throughout this paper we adopt a cosmological model with $\Omega = 1$, $\Lambda = 0$, and a "filled beam" approximation; the results, however, do not depend on the cosmology due to the low redshift of the lensing galaxy.

As the lens model we use an axially symmetric mass distribution plus external shear $\gamma$ as a simple and convenient representation of an astigmatic lens. The mass $m$ contained within radius $r$ is parametrized as

$$m(<r) = m_0 \times r^\beta, \quad 0 \leq \beta < 2.0, \quad \beta = const.$$

The values of the exponent $\beta = 0, 1, 2,$ correspond to a point mass, a singular isothermal sphere (SIS), and a constant surface mass density, respectively, with the last case not actually included among our models (the largest value for which we could find a model at all was $\beta = 1.93$ ). The mass normalization $m_0$ was chosen such that the mass is fixed at the Einstein radius ($\theta_{Einstein}$, see below). All our models are singular at the center, so the fifth image is absent.

We use the normalized lens equation (cf. Schneider, Ehlers, & Falco 1992) in the following form:

$$\vec{y} = \begin{pmatrix} 1 - \gamma \cos \Phi_\gamma & -\gamma \sin \Phi_\gamma \\ -\gamma \sin \Phi_\gamma & 1 + \gamma \cos \Phi_\gamma \end{pmatrix} \vec{x} - \vec{\alpha}(\vec{x}),$$

$$\text{where} \quad \vec{\alpha}(\vec{x}) = |\vec{x}|^{\beta-1} \frac{\vec{x}}{|\vec{x}|} \quad \text{and} \quad \vec{x} = \vec{x}_{ray} - \vec{x}_{lens}.$$

Here $\vec{x}$ and $\vec{y}$ are the positions in the image and source plane, respectively, $\gamma$ is the value of the external shear, $\Phi_\gamma$ is the angle of the shear relative to the East-North coordinate system used, e.g., by Crane et al. (1991), $\vec{\alpha}(\vec{x})$ is the deflection angle and $\beta$ is the mass index defined above.

In the two-dimensional parameter space: mass exponent – external shear we look for the "best" model in a grid of points $(\beta_i, \gamma_j)$. For fixed values of $\beta$ and $\gamma$ the "best" model is defined as one with the lowest value of $\chi^2$, which is given as

$$\chi^2 = \sum_{k=1}^{10} \frac{(x_{k,th} - x_{k,obs})^2}{\sigma_k^2},$$

where $x_{1,th}, \ldots x_{8,th}$ and $x_{1,obs}, \ldots x_{8,obs}$ are respectively the theoretical and the observed positions of the four images A,B,C,D, and $x_{9,th}$, $x_{10,th}$, $x_{9,obs}$, $x_{10,obs}$ are the positions of the galactic center; $\sigma_k^2$ are the variances of the observed positions $x_{k,obs}$. We adopt the observational accuracy as given in Crane et al. (1991) as the standard variation of all positions: $\sigma_k = 0.''005$. For each pair of values: ($\beta_i$, $\gamma_j$) we optimize the positions by minimizing $\chi^2$. In order to find the minimum for the $\chi^2$ we used a downhill-simplex method according to Press et al. (1986), their routine AMOEBA. The following model parameters were allowed to vary: the position of the source $(x_{1,s}, x_{2,s})$, the angle of the external shear $\Phi_\gamma$, and the overall scale factor $c_{scale}$ which fixes the mass scale, or equivalently the Einstein ring radius.

That means that there are ten observational constraints: the positions of the four images (8) and the position of the center of the galaxy (2). On the other hand there are eight parameters to fit, almost as many as there are constraints: the source position (2), the lens position (2) the magnitude and the direction of the shear (2), the slope of the mass as a function of radius (1) and the normalization of the enclosed mass(1).

We scanned the parameter space over $0.0 \leq \gamma_j \leq 0.3$, with grid size $\Delta\gamma = 0.002$, for the shear, and over $0.0 \leq \beta_i < 2.0$, with grid size $\Delta\beta = 0.02$ for the mass exponent. Once the "best" model for a pair of parameters is found, other image properties are determined: the absolute magnifications $\mu_i$ of the individual images, the relative time delays between the images $\Delta t_{ij}$, the surface mass densities $\kappa_i$ at the image positions, the local values of the shear $\gamma_i$ at the image positions, the Einstein radius $\theta_{Einstein}$, the mass inside the Einstein radius $m(\theta \leq \theta_{Einstein})$, and the position of the source relative to the center of the lens $(x_{1,s}, x_{2,s})$.

## 3. Results

A large number of models for Q2237+0305 was obtained following the procedure described in Chapter 2. We run four different model sequences: in the first sequence we used the positions of Crane et al. (1991) and fitted all ten model coordinates with equal weight. In the second sequence we chose the coordinates of the lens center to be free parameters, i.e. we fitted only the eight coordinates of the four images. In sequences three and four we did the same for the Racine (1991) positions. Below we shall give results only for sequence one, because the other three cases were only slightly different.

In Figure 1 the two-dimensional parameter space $\beta - \gamma$ is displayed. Each dot corresponds to a pair of values $(\beta_i, \gamma_j)$ for which we have calculated a model. The thick solid lines (labelled "68.3%" and "99.73%") are confidence region ellipses with values of $\chi^2$ larger than the minimum. The encircled regions contain 68.3% and 99.73% of normally distributed models around the minimum (cf. Press et al. 1986). From these contours it is, e.g., obvious that the best isothermal model is not significantly worse than the very best model. The location of the valley of best models is well described with the straight line $\gamma = 0.1385 - 0.0689\beta$.

For any value of the mass exponent $\beta$ (or the external shear $\gamma$) there is a clear minimum of $\chi^2$, i.e. the "best" model for this value of $\beta$. The best point mass model ($\beta = 0.0$) is obtained for a value of $\gamma = 0.138$ ($\chi^2 = 18.9$). Best isothermal sphere model ($\beta = 1.0$) for $\gamma = 0.070$ ($\chi^2 = 16.1$). The very best model in the whole plane with $\chi^2 = 14.2$ is found for $\beta = 1.710$ and $\gamma = 0.020$. However, the "best models" for any value of $\beta$ in the range $0 \leq \beta \leq 1.85$ are all very good, and they can hardly be distinguished from each other just by looking at the agreement between the observed and the model image positions. In fact the rms errors of the image positions for all these models are factors of 3 to 4 smaller than those given by R92.

Along the thin solid lines in Figure 1 the total magnification is constant. The value of the total magnification increases strongly along the "best model valley", from $\mu_{tot} \approx 8$ for $\beta = 0.0$ to $\mu_{tot} > 1000$ for $\beta > 1.85$.

The shear angle varies very little along the valley of best models, its value is always $\Phi_\gamma = (-23.16 \pm 0.03)°$. This direction corresponds to a position angle of $\theta = 90° + \Phi_\gamma = (66.84 \pm 0.03)°$ and agrees perfectly with the value for the model major axis found in KF88, $\theta = (66.8 \pm 0.5)°$. In Fig. 1 of KF88 this direction is shown to be between the position angle of the disk major axis (77°) and the position angle of the bar (38°). Our value is also very close to the position angle of 68° found by R92 and that of K91, who finds values between 65° and 67°.

In Figure 2 the image configuration is shown for the best isothermal sphere model, $\beta = 1.00$, $\gamma = 0.070$. The observed positions (Crane et al. 1991) of the four images are marked with small black dots, the model positions are indicated with small circles. The large dashed circle is the Einstein ring for this model. The "offset" for all four images in the SIS model displayed here is less than 0.01 arcsec per image. The offset of the center of the lens relative to the core of the galaxy is 0.013 arcsec – this one is always the largest, which prompted us to run sequences two and four with the position of the center of the lens not constrained at all. The resulting models were slightly better in terms of positional agreement between the four images than those described here, but not by a large margin.

The time delay lines for the best isothermal sphere model are also displayed in Figure 2. Real images occur at the stationary points of the time delay surface (cf. Schneider, Ehlers & Falco 1992). Each contour line corresponds to one hour. Images B and A are located at the minima of the time delay surface, whereas images C and D are are located at the saddle points. In this singular isothermal sphere model the time delay between images A and B is about one hour; note, however, that image B is leading. The time delay between images A and D is of order 3 hours, between A and C about 10 hours (for $H_0 = 75$ km sec$^{-1}$ Mpc$^{-1}$).

In Figure 3 different properties of the "best models" as a function of $\beta$ are displayed: The goodness of fit $\chi^2$ varies insignificantly along the valley of the best models. The agreement between the "best models" and the observations gets slightly better with increasing $\beta$, a broad minimum is reached at values of $\beta$ slightly larger than 1.0. For $\beta$ approaching 2.0 the agreement deteriorates considerably. In terms of image geometry the "best models" are almost indistinguishable from each other for $0.4 \leq \beta \leq 1.6$.

The value of external shear required by the "best models" decreases linearly from $\gamma \approx 0.14$ to 0.01 while $\beta$ increases from 0 to about 1.93, as can be seen in the second panel of Figure 3. The Einstein radius is almost independent of $\beta$: it increases from $\theta_{Einstein} \approx 0.869$ to 0.876 arcsec with increasing $\beta$ (third panel).

The total magnifications, i.e. the sum of the four individual magnifications: $\mu_{tot} = \sum_{i=1}^{4} |\mu_i|$, changes dramatically along the valley of good models, from $\mu_{tot} \approx 8$ for the point mass model ($\beta = 0$) to $\mu_{tot} > 1000$ for large $\beta$. The images C and D have negative parity as they are located at the saddle points of the time delay surface. Hence, their formal magnifications are negative. Throughout this paper we use the absolute values of all magnifications. The best isothermal model has the total magnification of $\mu_{tot} = 31.4$; and the very best model ($\beta = 1.71$) has $\mu_{tot} = 376.1$.

In the last panel of Figure 3 the total mass enclosed in an Einstein ring is shown as a function of $\beta$. We find that that mass, $m(\theta \leq \theta_{Einstein})$, is almost constant for all our models, varying between 1.475 and $1.493 \times 10^{10} M_\odot \, h_{75}^{-1}$.

In Figure 4 the relative magnifications $\mu_{BA}$ (solid line), $\mu_{CA}$ (dotted line) and $\mu_{DA}$ (dashed line) are plotted. The ratio between images B and A is almost constant at $\mu_{BA} \approx 0.85$. The ratio $\mu_{CA}$ increases slightly with $\beta$, from $\mu_{CB} \approx 0.5$ to 0.6. These two ratios roughly agree with the observations (cf. table 2 of Witt and Mao, 1993), The ratio $\mu_{DA}$ (dashed line) varies from about 1.0 to almost 1.2, and it is considerably higher than the observed value.

In the second panel of Figure 4 the values for the surface mass densities $\kappa_i$ at the positions of the four images are shown: A (solid line), B (dotted line), C (short-dashed line), and D (long-dashed line). The values of surface mass density increase from zero (for point lens) to about critical at large $\beta$. Notice, that for a given value of $\beta$ surface mass densities are similar for all images, and almost identical for images A and B, with $\kappa_C$ and $\kappa_D$ only slightly larger.

The local values of shear at the locations of the four images are shown in the third panel as decreasing functions of $\beta$. Again, $\gamma_A$ and $\gamma_B$ are almost identical, and somewhat smaller than $\gamma_C$ and $\gamma_D$.

The fourth (and last) panel shows the relative time delays $\Delta t_{Ai}$ between image A and the other three images along the valley of the best fit models. All numbers are given for the Hubble constant $H_0 = 75$ km sec$^{-1}$ Mpc$^{-1}$; naturally, $\Delta t \propto H_0^{-1}$. In all our models image B is leading: an intrinsic change in the quasar would show up first in image B, then (after a small time interval) in image A, then D, and the trailing image is always C. In other words: the maximum time delay for a given model always occurs between images B and C. This is in agreement with the sequence of S88 (see their Erratum) as well as with model 2a of R92, which is their best model and the only one that was constrained solely by the image positions; in all other R92 models the A image is leading. In our models the largest values of the time delays were obtained for the smallest $\beta$, i.e. for the point lens case, for which $\Delta t_{BC} \approx 20$ hours; with increasing $\beta$ this value goes down to time delays as small as $\Delta t_{BC} \approx 1.5$ hours (for $\beta \approx 1.85$). The relative time delays between images B-A and B-D scale as: $\Delta t_{BA} \approx 0.15 \times \Delta t_{BC}$, and $\Delta t_{BD} \approx 0.39 \times \Delta t_{BC}$. The dependence of the relative time delays on the parameter $\beta$ can be seen on the last panel in Figure 4: all three relative time delays decrease in the same way strongly with increasing $\beta$.

The parameters for three lens models are given in Table 1. The distance of the source from the galaxy center was equal 0.071, 0.036, and 0.010 arcsec for $\beta = 0$, 1.0, and 1.71, respectively.

## 4. Discussion and Conclusions

The total mass inside the Einstein ring is $(1.48 \pm 0.01) \times 10^{10} h_{75}^{-1} M_\odot$ for all our models, it varies very little with the value of $\beta$ (cf. Table 1 and Figure 3). For the lens model with constant mass-to-light ratio R92 had obtained a value of $1.08 \pm 0.02 \times 10^{10} h_{100}^{-1} M_\odot$ for the mass inside their Einstein radius of 0.9 arcsec. This corresponds to $m \approx 1.44 \pm 0.03 \times 10^{10} h_{75}^{-1} M_\odot$. That means there is perfect agreement between the mass estimates of the two quite different sets of models. This means that the mass depends very weakly on the lens model. As a consequence, this is probably the most accurate mass estimate for an extragalactic system.

Looking at the image configurations with respect to the Einstein ring in Figure 2 it is not too surprising that the mass obtained with various models is almost independent of the mass exponent $\beta$: the four images of Q2237+0305 are all very close to the Einstein ring. As all images are at about the same distance from the lens, it is not the slope of the mass distribution that is probed, but rather the total mass inside this radius. This result was first obtained by Kochanek (1991). He also finds that the values of the global shear are $\gamma \approx 0.14$ for his point mass models and around 0.07 for the isothermal models, very similar to what we get.

A macro model of a gravitational lens system usually yields values of the local convergence and shear at the position of the images. However, the value of the convergence or normalized surface mass density $\kappa_i = \kappa_{i,compact} + \kappa_{i,smooth}$ at the position of the quasar image #$i$ does not yet say anything about the frequency of microlensing events, because it is not clear what part of this matter is smoothly distributed ($\kappa_{i,smooth}$) and what part is in objects with masses of about $10^{-4} < m/M_\odot < 100$ ($\kappa_{i,compact}$). The (average) magnification will be the same, but the frequency and amplitude of any flucutation do depend on the nature of the matter. Now, as it is likely that in the lensing galaxy considered here about 50% or more of the mass in the inner 500pc of the lensing galaxy is in old bulge stars, and as the "effective" surface mass density scales as $\kappa_{eff} = \kappa_{compact}/(1 - \kappa_{smooth})$ (Paczyński 1986), a certain macro model with its values of $\kappa_i, \gamma_i$ for the images does constrain the microlensing properties of the lensed images.

Panels 2 and 3 in Figure 4 show that in our series of models the microlensing parameters

$\kappa_i$ and $\gamma_i$ of the four images can have a large range of values. The macro-image geometry does not constrain these numbers at all. In contrast, if one assumes that a large fraction of the matter in the innermost 500pc of the lensing galaxy is in the form of compact objects of roughly solar mass (i.e. low mass stars or brown dwarfs), then the observations of microlensing events can be used to constrain the models: the fact that microlensing does occur in this system at a rate of at least one event per year (Corrigan et al. 1991, Houde & Racine 1994) indicates that the actual values of the surface mass density must be larger than, say, $\kappa = 0.2$, otherwise one would expect fewer microlens-induced changes in the light curve. On the other hand, if $\kappa$ was close to 1.0, then one would expect many more microlensing events if the source is small, or a very smooth light curve, if the source is large and always covers many micro-caustics (cf. Wambsganss 1990). Therefore "reasonable" values of $\kappa$ are likely to be between 0.2 and 0.8.

It is surprising how well our very simple models reproduce the image positions. Our "very best" model ($\beta = 1.71$, $\gamma = 0.02$) has the total magnification $\mu = 376.1$, and the maximum time delay is $\Delta t_{BC} = 2.98$ hours. The rms positional agreement between the calculated and the observed positions of all four images are $\sim 0.007$ arcsec, a factor $\sim 4$ better than in the best R92 model. The positional accuracy is almost as good along all the "valley of good models" for $0.0 \leq \beta \leq 1.85$.

With our parametrization of lens astigmatism we found a one parameter family of solutions that reproduce the macro - image geometry far better than any other model to date. This indicates that the geometry alone is not sufficient to make a unique model (cf. K91). Some additional physical constraints are needed. To a good approximation the intensity ratios and the time delay ratios are about the same in all models that reproduce the geometry well, but the total magnification as well as the scaling for the time delay do vary along our family of solutions. It will be interesting to check if the same qualitative property holds for other parametrizations of the lens astigmatism (e.g. Kovner 1987)

The models based on the assumption of constant mass to light ratio, like R92, are conceptually simple, and provide a fairly accurate description of the macro - image geometry. However, as simple and attractive the assumption of constant M/L ratio is, it can only be an approximation to any real galaxy. Also, the reddening is known to vary across the lensing galaxy, but it is very difficult to determine it quantitatively. The differences between the best R92 model positions and the observed image positions are far larger than the measurement errors, which indicates that the constant M/L ratio, as adopted, is a fair but not very accurate approximation of the lens. In particular, at this time we do not know what is the range of total magnifications which are compatible with slight variations of the M/L ratio, variations which almost certainly must be present in every galaxy. In fact,

– 10 –

it should be possible to quantify the departure from a constant M/L ratio in the lensing galaxy. The good agreement of the purely geometrical models with the observations as presented here and by Kochanek (1991) suggests that physical models with variable M/L ratio could result in improved models, compared to R92. However, the experience with the degeneracy of the geometrical models shows that there is probably no unique way to model such a lens. Our simple models demonstrate that the range of model magnifications may be surprisingly large.

In principle there are ways to break the parameter degeneracy mentioned throughout this paper. One method would be to find the actual values of the surface mass density $\kappa_i$ at the positions of the four images from observations of microlensing. As indicated above, comparison of amplitudes and frequencies of observed microlens-induced changes (see Corrigan et al. 1991, or Pen et al. 1993) with simulations already narrow in on the range of possible values of the surface mass density. For tighter constraints on the $\kappa_i$'s, however, one has to make some assumptions on the exact size of the quasar, on the transverse velocity, and on the mass of the microlensing objects (not to mention the need for much better resolved observational light curves), so that one does not win very much.

A second way to break the parameter degeneracy could be the detection of a time delay in this system. For that, one needs very closely spaced observations ($\Delta t < 1$ hour), and, more importantly, the quasar must intrinsically vary on a time scale comparable to or shorter than the time delay. This is close to impossible in the optical continuum. In other wave bands, however, it is at least imaginable. A highly Doppler boosted radio jet can certainly vary on short time scales. Maybe even more interesting is the possibility of a short variation, flash-like, in the X-ray regime. There is not very much known about the X-ray and radio properties of the quasar 2237+0305; in order to learn more about the lensing galaxy, this may be worthwhile exploring.

On the other hand, the dependence of the time delay on the mass profile shows that it may turn out to be difficult to determine the Hubble constant from a measurment of the time delay, at least in this system. As the slope of the time delay $\Delta t$ with the mass index $\beta$ is as steep as $d \log \Delta t/d\beta \approx 1$, it seems unlikely that an accurate value of $H_0$ can be obtained from this system, unless there are quite accurate independent measurements of the mass distribution for the lensing galaxy, that narrow in the allowed range of $\beta$. The influence of this parameter degeneracy on the time delay in other lens systems, and in particular its consequences for determining the Hubble constant, deserves further investigation.

It is a pleasure to thank Jürgen Ehlers, Emilio Falco, Shude Mao, Paul Schechter, Peter Schneider, Hans-Jörg Witt, and especially Hans-Walter Rix for many useful discussions and comments, and to acknowledge NATO Collaborative Research Grant No. 900209 and the NASA grant NAGW-765.

---

Table 1: Parameters for three models of 2237+0305: point mass lens, singular isothermal sphere, and the best-fit model: $\beta$ – mass index; $\gamma$ – external shear; $\chi^2$ – goodness-of-fit; $\mu_{tot}$ – total magnification; $\mu_{BA}$, $\mu_{CA}$, $\mu_{DA}$ – magnification ratios between images B-A, C-A, and D-A; $\Delta\tau_{AB}$, $\Delta\tau_{AC}$, $\Delta\tau_{AD}$ – relative time delays between images A-B, A-C, and A-D (in hours); $\theta_E$ – Einstein radius (in arcsec); $M_{\theta<\theta_E}$ – mass inside Einstein ring (in $10^{10}$ $M_\odot$); $\kappa_i$ – (dimensionless) surface mass density of this image; $\gamma_i$ – local value of shear at position of this image; $\Delta_i$ – difference between observed and model position of image (in arcsec).

| Model | point lens | isothermal | best fit |
|---|---|---|---|
| $\beta$ | 0.00 | 1.00 | 1.71 |
| $\gamma$ | 0.138 | 0.07 | 0.02 |
| $\chi^2$ | 18.98 | 16.1 | 14.2 |
| $\mu_{tot}$ | 8.10 | 31.4 | 376.1 |
| $\mu_{BA}$ | 0.876 | 0.859 | 0.847 |
| $-\mu_{CA}$ | 0.479 | 0.562 | 0.637 |
| $-\mu_{DA}$ | 0.978 | 1.073 | 1.144 |
| $\Delta\tau_{AB}$/hours | 2.97 | 1.51 | 0.44 |
| $\Delta\tau_{CB}$/hours | 20.38 | 10.42 | 2.98 |
| $\Delta\tau_{DB}$/hours | 7.84 | 3.97 | 1.14 |
| $\theta_E$/arcsec | 0.870 | 0.873 | 0.874 |
| $M_{\theta<\theta_E}/10^{10}\ M_\odot$ | 1.475 | 1.484 | 1.490 |
| $\kappa_A$ | 0.000 | 0.469 | 0.840 |
| $\kappa_B$ | 0.000 | 0.461 | 0.836 |
| $\kappa_C$ | 0.000 | 0.558 | 0.883 |
| $\kappa_D$ | 0.000 | 0.514 | 0.862 |
| $\gamma_A$ | 0.767 | 0.413 | 0.126 |
| $\gamma_B$ | 0.728 | 0.401 | 0.125 |
| $\gamma_C$ | 1.363 | 0.627 | 0.170 |
| $\gamma_D$ | 1.192 | 0.583 | 0.166 |
| $\Delta_A$/arcsec | 0.004 | 0.003 | 0.004 |
| $\Delta_B$/arcsec | 0.009 | 0.009 | 0.008 |
| $\Delta_C$/arcsec | 0.010 | 0.009 | 0.009 |
| $\Delta_D$/arcsec | 0.009 | 0.009 | 0.008 |
| $\Delta_{gal}$/arcsec | 0.016 | 0.014 | 0.013 |

– 15 –Fig. 1.— Two-dimensional parameter space of the mass index $\beta$ and the external shear $\gamma$. Each point indicates one pair $(\beta_i, \gamma_j)$. The thick solid lines connect points with the same $\Delta\chi^2$, relative to the minimum. These confidence region ellipses contain 68.3% and 99.7% of normally distributed models around the minimum of $\chi^2$. The thin solid lines connect models with equal total magnification $\mu_{tot}$.

Fig. 2.— Image configuration and time delay lines for the best isothermal sphere model ($\beta = 1.0$, $\gamma = 0.07$) of Q2237+0305. The points mark the observed positions of the quasar images and the galaxy core, the circles indicate the positions of the four model images and the center of the lens. The large dashed circle is the Einstein ring for this configuration. Its radius is $\theta_{Einstein} \approx 0.869$ arcsec. The contour lines for the time delay are spaced by one hour. The images A and B are located at the minima of the time delay, and the images C and D are located in the saddle points. The scale is in arcseconds. In the lower right corner the letters A-D indicate the labelling of the images.

Fig. 3.— Dependence of lens parameters as a function of $\beta$. From top to bottom: the goodness of fit $\chi^2$, the external shear $\gamma$, the radius of Einstein ring $\theta_{Einstein}$ in arcsec, the logarithm of total magnification $\log \mu_{tot}$, the total mass inside Einstein ring $m(< \theta_{Einstein})$ in units of $10^{10} M_\odot h_{75}^{-1}$.

Fig. 4.— The variation of image parameters with $\beta$. From top to bottom: The magnification ratios $\mu_{BA}$ (solid), $\mu_{CA}$ (dotted), $\mu_{DA}$ (dashed); the values of surface mass density $\kappa_i$ at the location of the four images (A: solid line; B: dotted; C: short-dashed; D: long-dashed); the values of local shear $\gamma_i$ at the location of the four images with the same line convention as for $\kappa$); and the relative time delays $\Delta t_{BA}$ (solid), $\Delta t_{CA}$ (dotted), and $\Delta t_{DA}$ (dashed) in hours. The dash-dotted line indicates a time delay of zero, to emphasize the negative value of $\Delta t_{BA}$, which means that image B leads image A.